\documentclass[preprint,3p,12pt]{elsarticle}
\usepackage{amsfonts}
\usepackage{epsf}
\usepackage{colordvi}
\usepackage{color}
\newcommand{\nwl}{\\begin{equation}2mm]}
\newcommand{\edc}{\end{document}}
\newcommand{\bb} {\begin{thebibliography}}
\newcommand{\eb}{\end{thebibliography}}
\newcommand{\bi}[1]{\bibitem{#1}}
\newcommand{\bc}{\begin{center}}
\newcommand{\ec}{\end{center}}

\newcommand{\be}{\begin{equation}\small}
\newcommand{\ee}{\end{equation}\normalsize}
\newcommand{\bea}{\begin{eqnarray}}
\newcommand{\eea}{\end{eqnarray}}
\newcommand{\ba}{\begin{array}{l}   }
\newcommand{\lab}[1]{\label{#1}}
\newcommand{\ea}{\end{array}}

\newcommand{\dsfrac}{\displaystyle\frac}
\newcommand{\ds} {\displaystyle}

\newcommand{\re}[1]{(\ref{#1})}

\newcommand{\ci}{\cite}

\newcommand{\dsint}{\ds\int}

\newcommand{\mathbfr}{ {\mathbf r}}

\newcommand{\eps}{\varepsilon}

\newcommand{{\vergul}}{  ,}

\newcommand{\veps}{\varepsilon }

\journal{Annals of Physics}

\begin{document}

\begin{frontmatter}

\title{Macroscopic properties of triplon Bose-Einstein condensates}

\author[Tashkent]{Abdulla Rakhimov}
\author[Samarkand]{Shuhrat Mardonov}
\author[Bilbao,Ikerbasque] {E. Ya. Sherman\corref{cor1}}
\cortext[cor1]{Corresponding author}
\address[Tashkent]{Institute of Nuclear Physics, Tashkent 100214, Uzbekistan }
\address[Samarkand]{Samarkand State University, Samarkand, Uzbekistan}
\address[Bilbao]{ Department of Physical Chemistry, Universidad del Pa\'is Vasco UPV/EHU,
48080 Bilbao, Spain }
\address[Ikerbasque] {IKERBASQUE, Basque Foundation for Science, 48011, Bilbao, Spain }

\begin{abstract}
Magnetic insulators can be characterized by a gap separating  
the singlet ground state from the lowest energy triplet, $S=1$ excitation.
If the gap can be closed by the Zeeman interaction in applied magnetic field, 
the resulting $S=1$ quasiparticles, triplons, can have concentrations
sufficient to undergo the Bose-Einstein condensates transition.  
We consider macroscopic properties of the triplon  Bose-Einstein condensates
in the Hartree-Fock-Bogoliubov approximation
taking into account the anomalous averages. We prove that these averages play the 
qualitative role in the condensate properties. As a result, 
we  show that with the increase in the external magnetic
field at a given temperature, the condensate 
demonstrates an instability related to the appearance of 
nonzero phonon damping and a change in the
characteristic dependence of the speed of sound on the magnetic field. 
The calculated magnetic susceptibility diverges
when the  external magnetic field approaches this instability threshold,
providing a tool for the experimental verification of this approach.
\end{abstract}

\begin{keyword}
{Bose condensation, specific heat, magnetic susceptibility} 
\PACS{75.45+j, 03.75.Hh, 75.30.D}
\end{keyword}

\end{frontmatter}

\section{Introduction}

\begin{figure}
\centerline{\includegraphics*[scale=0.6]{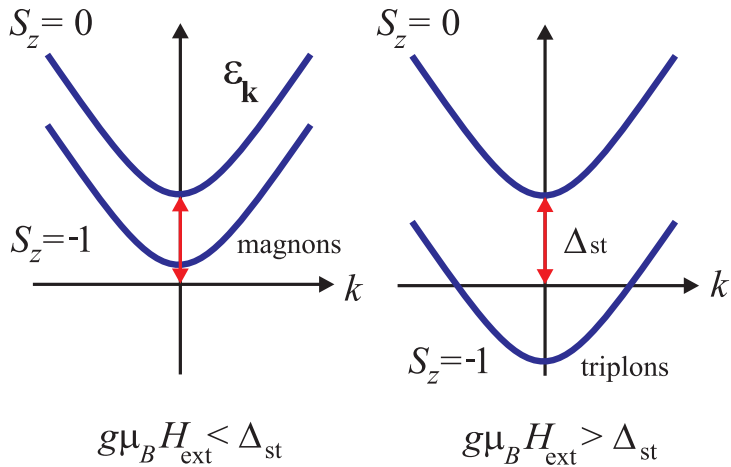}}
\caption{(Color online) Evolution of the magnon states with the increase in the magnetic field.
When the bottom of $S_z=-1$ band has negative energy, magnons (left panel) become ``triplons'' (right panel).}
\label{triplons}
\end{figure}

Macroscopic systems governed by quantum mechanics of interacting particles, with ensembles 
of cold atoms and quantum magnets being the intensively 
investigated examples, attract a great deal of interest. As well understood, 
they have interesting similarities, showing the common physics of these two
seemingly different realizations. Macroscopic ensembles of bosonic atoms can
undergo Bose-Einstein condensation (BEC). The properties of these atomic 
condensates in different systems became one of the most interesting fields
in modern experimental and theoretical physics. 
On the other hand, triplet quasiparticles in  magnetic insulators, 
being bosons with spin equal to one, can undergo the BEC transition.
The first example of this kind of condensate 
is the systems far away from the equilibrium,
where high concentration of spin excitations (magnons) is achieved by a strong resonant external
magnetic field pumping \cite{Matsuda,Tachiki,Batyev,Demidov08}. At a sufficiently 
intense pumping, this type of condensation can occur at temperatures of the order of 100 K. 
Second example \cite{Volovik} is the magnons
in superfluid He$^{3}$, where the magnetism appears due to the very small nuclear rather than electron magnetic moment.
This process occurs at temperatures lower than 10$^{-3}$ K.  
Third interesting example is presented by triplons, appearing
if the gap in the triplet excitations spectrum 
separating it from the singlet ground state,
can be closed by the Zeeman effect of the magnetic field \cite{Affleck,Giamarchi99}. 
The field splits the spin $S=1$ excitations into three branches with
$S_z=0,\pm1$. When the Zeeman shift of the $S_z=-1$ branch exceeds the initial gap
in the spectrum, a finite population of triplons, which can be considered as the
reconstruction of the ground state, is formed even at zero temperature, as shown in Fig.\ref{triplons}. 
The BEC transition is seen as the transverse magnetization,
which occurs if the field becomes higher than the critical one \cite{Giamarchi08}. 
The Bose-Einstein condensation of triplons was experimentally first observed \cite{Nikuni00,oosawa} 
and thoroughly studied in the magnetic insulator TlCuCl$_{3}$ \cite{oosawa} where magnetic properties
are due to the presence of noncompensated single-electron spins of Cu ions. Similar findings
in a variety of other compounds followed shortly \cite{Ruegg07,Stone07,Aczel09,Paduan09,Laflorencie09}. 
Recently, Bose-Einstein condensation of magnetic excitations was studied theoretically
in compounds, where the magnetism is due to the spins of vanadium \cite{Tsirlin} or chromium \cite{Kim} ions.
These uniform, three dimensional with various degrees of anisotropy and easily tunable
systems provide the researchers with new abilities to study the BEC, including the 
effects of disorder. The tunability is realized by 
easily changing the chemical potential with the external magnetic field $H_{\rm ext}$.
The condensates were studied in a variety of magnetic fields 
up to the fields where the Zeeman effect strongly
changes the properties of the systems due to the magnetization saturation. 

Since the triplon BEC occurs in solids, it possesses at least three interesting features
resulting from its coupling to the host lattice. 
First, it is directly influenced by the spin-orbit coupling inherent to the solid where
it is located \cite{Sirker04}. For atomic BEC the analog of spin-orbit coupling can be produced 
by special combination of laser fields in some cases \cite{Stanescu08}. Second, the triplon BEC is coupled
to phonons, and, therefore, can provide a test for the effects of decoherence and hysteresis due to the coupling
to the lattice \cite{Sherman03,Continentino}. Third, 
an external pressure applied to the crystal can strongly change the properties
of the triplon BEC \cite{Oosawa04,Kulik11}.   In addition, the solid-state background for the triplon
BEC makes it accessible by a variety of experimental techniques, not applicable
for the atomic condensates. 

Theoretical description of triplon BEC in relatively weak fields, where
the Zeeman splitting is much less than the width of the triplon band in the Brillouin
zone, and, correspondingly, concentration of triplons is small, can be done
in terms of weakly interacting Bose gas theory. The observation of the sound-like Bogoliubov spin-flip mode,
the fingerprint of the BEC in an interacting Bose gas, 
in the neutron scattering experiment \cite{Ruegg} provided the strong support
of this approach to the BEC of triplons.
 
Main tools for these studies 
are based on the approach usually referred to  as the Hartree-Fock-Popov (HFP) approximation, neglecting
the anomalous density 
terms\footnote{Although connecting this approximation with the name of Popov in the 
literature is not adequate since Popov 
never neglected $\sigma$-terms (V.N. Popov, Functional Integrals in 
Quantum Field Theory and Statistical Physics, Reidel, Dordrecht, 1983) 
we will use the acronym HFP approximation for the historical reasons.}. 
The drawback of this approximation  is the prediction of a 
jump in the triplon density, and, in turn, in the sample magnetization across the transition
temperature. Although this approach provides 
a good quantitative description of the specific heat $C_{v}$, it is unclear
whether it captures qualitatively all relevant
physics of the condensate. For this reason it will be interesting 
to study the behavior of the triplon BEC using a more accurate approach in order
to see whether the analysis beyond the HFP approximation reveals new qualitative
features.

In our previous work \ci{ourprb} we have shown that an extended version of the  mean field
approach (MFA) taking into account anomalous density terms in the condensate,
improves the situation considerably \cite{Crisan05}. 
The key part of this approach concentrates on finding the speed of 
sound-like Bogoliubov excitations in the BEC of interacting particles. 
This Hartree-Fock-Bogoliubov (HFB) approximation \ci{yukANN}  
leads to a continuous magnetization across the
transition, in agreement with the experiment. 
By applying this approximation to the BEC of triplons
we have shown that when the external magnetic field $H_{\rm ext}$ exceeds a critical value,
$H^{\rm cr}_{\rm ext}$, the speed of sound becomes complex and the BEC state undergoes an instability.

Another general feature clearly seen in the triplon BEC is the dependence of its physics
on the bare magnon dispersion $\varepsilon_{\mathbf k}$. The non-parabolic bare dispersion
of magnons \cite{Sherman03,Misguich04,Jensen,Keeling} leads to a non-trivial dependence
of the transition temperature $T_{\rm c}$ on the concentration
$\rho\sim M(T,H_{\rm ext})$ and, hence, on $H_{\rm ext}$.
The bare dispersion, being itself $H_{\rm ext}$-independent,
determines the interplay of kinetic and potential energy
of a macroscopic system, and, therefore, plays a crucial
role in the BEC properties.
The effects of the bare dispersion are clearly seen experimentally in TlCuCl$_3$
as the $\rho$-dependence $T_{\rm c}\sim\rho^{\phi(\rho)}$.
The exponent ${\phi(\rho)}$ approaches $2/3$ at low concentrations (low $T_c$),\ci{Yamada08}
as expected for the parabolic $\varepsilon_{k}\sim k^{2}$, while
at $T>2.5$ K, $\phi(\rho)$ is close to 0.5. We will address the role of the dispersion
below in the paper and present simplified for the parabolic dispersion calculations.

In this paper we address the macroscopic properties of the BEC in terms of the speed of sound,
specific heat, and magnetic susceptibility, in the HFB approximation
and show both the quantitative and qualitative importance of the anomalous averages. 
This paper is organized in the following way. In Section II we outline main
features of the HFB approximation. In Section III we discuss the specific 
heat and magnetic susceptibility in this approximation and show the importance of the
anomalous averages. In Section IV we provide a detailed analysis of the instability caused
by external magnetic field in terms of the speed of sound and magnetization. 
This instability is the qualitative effect, not expected
if the anomalous averages are not taken into account. The Conclusions summarize our results. 

\section{The Hartree-Fock-Bogoliubov approximation.}

We begin with the Hamiltonian of triplons as a nonideal Bose gas with
contact repulsive interaction:
\be
H=\int d^{3}r 
\left[
\psi^{\dagger}(\mathbf{r})\left(\hat{K}-\mu\right)\psi(\mathbf{r})+
\frac{U}{2}\left(\psi^{\dagger}(\mathbf{r})\psi(\mathbf{r})\right)^{2}
\right],
\label{2.1}
\ee
where $\psi(\mathbf{r})$ is the bosonic field operator, $U$ is the 
interaction strength, and $\hat{K}$ is the kinetic energy operator which
defines the bare  triplon dispersion $\varepsilon_{\mathbf k}$ in momentum space. 
Since the triplon BEC occurs in solids, 
we perform integration over the unit cell of the crystal with the corresponding 
momenta defined in the first Brillouin zone. Below the spectrum
will be assumed isotropic: $\varepsilon_{\mathbf k}=\varepsilon(k)$. The parameter $\mu$ characterizes
an additional direct contribution to the triplon energy due to the external field and  being rewritten as
\be
\mu=g\mu_{B}H_{\rm ext}-\Delta_{\rm st}
\lab{2.2}
\ee
can be interpreted as a chemical potential of the $S_z=-1$ triplons \ci{Giamarchi08}.
In Eq. \re{2.2} $g$ is the electron Land\'e factor, $\mu_{B}$ is the Bohr magneton and $\Delta_{\rm st}$ is the spin gap
separating the singlet ground state from the lowest-energy triplet excitation (Fig.\ref{triplons}). 

In general the Hamiltonian in \re{2.1} is invariant under global $U(1)$ gauge transformation
\be
\psi({\mathbf r})\rightarrow e^{i\alpha}\psi({\mathbf r})
\label{U1}
\ee
with $\alpha$ being a real number. However, this symmetry is broken in the condensed phase, where $T\leq T_{c}$,
and restored for the normal phase,  $T>T_{c}$. Note that in the experimental
studies of BEC of atomic gases the density $\rho$ 
(or equivalently, the total number of atoms) is fixed
initially and the chemical potential $\mu$ should be calculated self-consistently,
while in the case of triplon BEC the chemical potential is fixed by the magnetic field and the density
should be determined within an appropriate approximation.

Here and below we adopt the units $k_{B}\equiv1$ for the Boltzmann constant, $\hbar\equiv1$ 
for the Planck constant, and $V\equiv1$ for the unit cell volume.
In these units the energies are
measured in Kelvin, the mass $m$ is expressed in K$^{-1}$, the magnetic susceptibility
$\chi$ for the magnetic fields measured in Tesla has the units of K/T$^2$, while the momentum 
and specific heat $C_v$ are dimensionless. Particularly, the Bohr magneton is 
$\mu_{B}={\hbar e}/{2m_{0}c}=0.671668$ K/T, where $m_{0}$ is the free electron mass, and $e$ is the fundamental charge.

\subsection {Condensate phase $T\leq T_{c}$}

As it has been  shown by Yukalov  \ci{yukANN} the spontaneous gauge symmetry breaking
is the necessary and sufficient condition for Bose-Einstein condensation and can be
realized by the Bogoliubov shift in the field operator as
\be
\psi(\mathbfr)=v(\mathbfr)+\widetilde{\psi}(\mathbfr),
\lab{2.3}
\ee
where for the uniform case the function $v(\mathbfr)$ is a constant defining the density of
condensed particles as
\be
\rho_{0}=v^{2}
\lab{2.4}
\ee
Since by the definition of the average of $\psi^{\dagger}(\mathbfr)\psi(\mathbfr)$ is 
the total number of particles:
\be
N=\int_{V}d^{3}r \langle\psi^{\dagger}(\mathbf{r})\psi(\mathbf{r})\rangle.
\lab{2.5}
\ee
with the density of triplons per unit cell $\rho=N/V$, from normalization  condition
\be
\rho=\rho_{0}+\rho_{1}
\lab{2.6}
\ee
one immediately obtains
\be
\rho_{1}=\dsfrac{1}{V}\int_{V}d^{3}r \langle\widetilde{\psi}^{\dagger}(\mathbf{r})\widetilde{\psi}(\mathbf{r})\rangle.
\lab{2.7}
\ee
Therefore  the field operator $\widetilde{\psi}(\mathbf{r})$ determines the density of uncondensed
particles.
Note that $\widetilde{\psi}(\mathbf{r})$ and the condensate function should be
considered  as independent variables being orthogonal to each other in terms of the expectation values:
\be
\int d^{3}r\widetilde{\psi}(\mathbfr)v(\mathbfr)=0.
\lab{2.7.1}
\ee
Consequently, passing to the momentum space
\be
\ds \widetilde{\psi}(\mathbf{r})=\sum_{\mathbf k}a_{\mathbf k}e^{i\mathbf{k}\mathbf{r}}
\equiv\int\frac{d^3{k}}{(2\pi)^{3}}a_{\mathbf k}e^{i\mathbf{k}\mathbf{r}}
\lab{2.9}
\ee
and inserting \re{2.9} into \re{2.7} one can easily see that in \re{2.9} the summation by momentum in
the finite volume should not include  $\mathbf{k=0}$ states. Below we indicate this rule by
introducing prime sign in the momentum summation.

Now using \re{2.3} and \re{2.9} in \re{2.1} we present the Hamiltonian as the sum of five terms
\bea
\ds H=\sum_{n=0}^{4}H_{n},
\lab{2.10}
\eea
labeled according to their order with respect to $a_{\mathbf k}$ and $a^{\dagger}_{\mathbf k}$.
The zero-order term does not contain the field operators of uncondensed triplons
\bea
\ds H_{0}=\frac{U}{2}\rho_{0}^{2}-\mu\rho_{0}.
\lab{2.11}
\eea
The first order term is identically zero
\be
H_{1}=0,
\lab{2.12}
\ee
due to the orthogonality condition \re{2.7}.
The second-order term is
\bea
\ds H_{2}=\sum_{\mathbf k}^{\prime}
\left[
\left(\varepsilon_{k}-\mu+2U\rho_{0}\right)a_{\mathbf k}^{\dagger}a_{\mathbf k}+
\frac{U\rho_{0}}{2}
\left(a_{\mathbf k}a_{-{\mathbf k}}+a_{\mathbf k}^{\dagger}a_{-{\mathbf k}}^{\dagger}\right)
\right].
\lab{2.13}
\eea
The third-order
\bea
\ds H_{3}=U\sqrt{\rho_{0}}\sum_{\mathbf {k,p}}
\left(a_{\mathbf p}^{\dagger}a_{\mathbf {p-k}}a_{\mathbf k}
+a_{\mathbf k}^{\dagger}a_{\mathbf {p-k}}^{\dagger}a_{\mathbf p}\right)
\lab{2.14}
\eea
as well as  fourth-order
\bea
\ds H_{4}=\frac{U}{2}\sum_{\mathbf {k,p,q}}'a_{\mathbf k}^{\dagger}a_{\mathbf p}^{\dagger}a_{\mathbf q}a_{\mathbf {k+p-q}}
\eea
 terms are  rather complicated  and a diagonalization procedure is  needed. For this purpose
in HFB approximation the following procedure is usually  implemented \cite{prokjack}:
\begin{eqnarray}
&&a_{\mathbf k}^{\dagger}a_{\mathbf p}a_{\mathbf q}\rightarrow2\langle a_{\mathbf k}^{\dagger} a_{\mathbf p}\rangle a_{\mathbf q}+a_{\mathbf k}^{\dagger}\langle a_{\mathbf p} a_{\mathbf p}\rangle,
\lab{2.14.1}\\
&&a_{\mathbf k}^{\dagger}a_{\mathbf p}^{\dagger}a_{\mathbf q}a_{\mathbf m}\rightarrow 4a_{\mathbf k}^{\dagger}a_{\mathbf m}\langle a_{\mathbf p}^{\dagger}a_{\mathbf q}\rangle
+a_{\mathbf q}a_{\mathbf m}\langle a_{\mathbf k}^{\dagger}a_{\mathbf p}^{\dagger}\rangle +a_{\mathbf k}^{\dagger}a_{\mathbf p}^{\dagger}\langle a_{\mathbf q}a_{\mathbf m}\rangle -2\rho_{1}^{2}-\sigma^{2},
\lab{2.15}
\end{eqnarray}
where
$\langle a_{\mathbf k}^{\dagger}a_{\mathbf p}\rangle=\delta_{\mathbf{k},\mathbf{p}}n_{\mathbf k},\quad \langle a_{\mathbf k}a_{\mathbf p}\rangle=\delta_{\mathbf{k},-\mathbf{p}}\sigma_{\mathbf k}$
with $n_{\mathbf k}$ and $\sigma_{\mathbf k}$ being related to the normal ($\rho_{1}$) and anomalous ($\sigma$) densities as
\bea
\ds \rho_{1}=\sum_{\mathbf k}n_{\mathbf k}, \qquad
\ds \sigma=\sum_{\mathbf k}\sigma_{\mathbf k}.
\lab{2.17}
\eea

Here we underline that the main difference between the HFP and HFB approximations concerns 
the anomalous density: neglecting
$\sigma$ as well as $\langle a_{\mathbf k} a_{\mathbf p}\rangle$ in \re{2.13}, \re{2.14.1}, and \re{2.15} one 
arrives at the HFP approximation,
which can also be obtained in variational perturbation theory \ci{Kleinert05}. 
However, the normal, $\rho_{1}$, and anomalous averages, $\sigma$, are equally important and 
neither of them can be neglected without making the theory not self-consistent \ci{AndersenRevModPhys,yukanom,Yukalov11}.
Altough $\rho_{1}$ and $\sigma$  are functions of temperature and
external magnetic field, we omit  explicit dependences in the formulas
when it does not cause a confusion. 

Taking into account that $E_{3}=\langle  H_{3}\rangle=0$, the expectation value 
of the energy can be evaluated as $E(T\leq T_{c})=E_{0}+E_{2}$, where 
\be
E_{0}=\dsfrac{U\rho_{0}^{2}}{2}-\mu\rho_{0}-\dsfrac{U}{2}(2\rho_{1}^{2}+\sigma^{2}),
\lab{2.18}
\ee
and $E_{2}=\langle\widetilde{H}_{2}\rangle$, with $\widetilde{H}_{2}$ being quadratic in $a_{\mathbf k}, a_{\mathbf k}^{\dagger}$:
\bea
\ds \widetilde{H}_{2}= \sum_{\mathbf k}\left[\omega_{k}a_{\mathbf k}^{\dagger}a_{\mathbf k}+
\frac{\Delta}{2}\left(a_{\mathbf k}a_{-{\mathbf k}}+a_{\mathbf k}^{\dagger}a_{-{\mathbf k}}^{\dagger}\right)\right].
\lab{2.19}
\eea
In the last equation
\bea
\omega_{k}=\varepsilon_{k}-\mu_{\rm eff}, \quad
\mu_{\rm eff}=\mu-2U\rho,
\lab{2.20}
\eea
\bea
\Delta=U(\rho_{0}+\sigma).
\lab{2.21}
\eea

The next step, in both approximations, is the Bogoliubov transformation
\bea
a_{\mathbf k}=u_{\mathbf k}b_{\mathbf k}+v_{\mathbf k}b_{-{\mathbf k}}^{\dagger}, \quad
a_{\mathbf k}^{\dagger}=u_{\mathbf k}b_{\mathbf k}^{\dagger}+v_{\mathbf k}b_{-{\mathbf k}}
\lab{2.22}
\eea
to diagonalize $\widetilde{H}_{2}$.
The operators $b_{\mathbf k}$ and $b_{\mathbf k}^{\dagger}$ can be interpreted as annihilation and creation operators of phonons with following properties:
\begin{eqnarray}
&&[b_{\mathbf k},b_{\mathbf p}^{\dagger}]=\delta_{\mathbf{k},\mathbf{p}},\quad  \langle b_{\mathbf k}^{\dagger}b_{-{\mathbf k}}^{\dagger}\rangle=\langle b_{\mathbf k}b_{-\mathbf{k}}\rangle=0,\\
&&\langle b_{\mathbf k}^{\dagger}b_{\mathbf k}\rangle=f_{B}(E_{k})=\dsfrac{1}{e^{\beta E_{k}}-1},
\lab{2.23}
\end{eqnarray}
where $\beta\equiv 1/T$. To determine the phonon dispersion $E_{k}$ we insert \re{2.22} into \re{2.19} and require that the coefficient
of the term $b_{\mathbf k}b_{-\mathbf{k}}+b_{-\mathbf{k}}^{\dagger}b_{\mathbf k}^{\dagger}$ vanishes, i.e:
\bea
\omega_{k}u_{\mathbf k}v_{\mathbf k}+\frac{\Delta}{2}\left(u^{2}_{\mathbf k}+v_{\mathbf k}^{2}\right)=0.
\lab{2.24}
\eea
Now using the condition $u_{\mathbf k}^{2}-v_{\mathbf k}^{2}=1$ and presenting $u_{\mathbf k}, v_{\mathbf k}$ as
\bea
u_{\mathbf k}^{2}=\frac{\omega_{k}+E_{k}}{2E_{k}},\quad \quad v_{\mathbf k}^{2}=\frac{\omega_{k}-E_{k}}{2E_{k}}
\lab{2.25}
\eea
yields
\bea
\sqrt{\omega_{k}^{2}-E_{k}^{2}}=-\Delta
\eea
that is
\bea
E_{k}=\sqrt{(\omega_{k}+\Delta)(\omega_{k}-\Delta)}
\lab{2.27}
\eea
where $\omega_{k}$ and $\Delta$ are given in Eqs. \re{2.20} and \re{2.21}.

Further requirement for $E_{k}$ follows from the Hugenholtz-Pines theorem \ci{stoofbook}:  at small momentum $k$ the spectrum should 
be gapless, and, therefore, the phonon dispersion is linear: $E_{k}\sim ck+O(k^{2})$,
where $c$ can be considered as the sound speed\footnote{It can be shown that \ci{MYPRA,AndersenRevModPhys} $\Delta$ is related to the normal ($\Sigma_{\rm n}$) and anomalous ($\Sigma_{\rm a}$) 
self-energies as $\Sigma_{\rm n}=\Delta+\mu$ and $\Sigma_{\rm a}=\Delta$, respectively.
The last two equations give $\Sigma_{\rm n}-\Sigma_{\rm a}=\mu$ which is again 
in agreement with Hugenholtz-Pines theorem \ci{Dickhoffbook,booknikinu}}.
This linearity can be achieved by setting
 \be
 \omega_{k}-\Delta=\varepsilon_{k},
 \lab{2.27.1}
 \ee
  which together with Eq.\re{2.21} yields
\be
\mu_{\rm eff}=\mu-2U\rho=-\Delta.
\lab{2.27.2}
\ee
With this choice one obtains
\be
E_{k}=\sqrt{\varepsilon_{k}}\sqrt{\varepsilon_{k}+2\Delta},
\ee
with the sound speed
\be
c=\sqrt{\frac{\Delta}{m}},
\lab{2.27.3}
\ee
i.e. $\Delta=m c^{2}$. Here $m$ has the meaning of the triplon effective mass, characterizing the dispersion
in the limit of small momenta $\varepsilon_{k}\approx k^{2}/2m$.

\subsection{Condensed fraction and the condensate energy.}

Having fixed $u_{\mathbf k}$ and $v_{\mathbf k}$ one can find normal and anomalous densities as well as the energy 
by using equations \re{2.17}-\re{2.25}. For example, inserting \re{2.22} into \re{2.17} results in
\bea
\ds \sigma=\sum_{\mathbf k}\langle a_{\mathbf k}a_{-{\mathbf k}}\rangle=
\sum_{\mathbf k}u_{\mathbf k}v_{\mathbf k}(1+2f_{B}(E_{k}))=
-\Delta\sum_{\mathbf k}\dsfrac{W_{k}}{E_{k}},
\lab{2.31}
\eea
where we used the relation $u_{\mathbf k}v_{\mathbf k}=-{\Delta}/{2E_{k}}$ and introduced notation $W_{k}={1}/{2}+f_{B}(E_{k})$.
Similarly one obtains:
\begin{eqnarray}
&&\ds\rho_{1}=\sum_{\mathbf k}\langle a_{\mathbf k}^{\dagger}a_{\mathbf k}\rangle=
\sum_{\mathbf k}\left(\dsfrac{W_{k}\omega_{k}}{E_{k}}-\dsfrac{1}{2}\right),
\lab{2.32}\\
&& E_{2}=\langle\widetilde{H}_{2}\rangle=\sum_{\mathbf k}E_{k}f_{B}(E_{k})+\dsfrac{1}{2}\sum_{\mathbf k}(E_{k}-\omega_{k}),
\lab{2.33} 
\end{eqnarray}
with  $\omega_k=\veps_k+\Delta$. 

When the bare dispersion $\veps_{k}$ is isotropic, the momentum summation can be done with:
\be
\sum_{\mathbf{k}}f\left({k}^{2}\right) =\dsfrac{4\pi }{\left( 2\pi \right) ^{3}}%
\int_{0}^{\infty} f\left({k}^{2}\right) k^{2}dk.
\lab{2.35}
\ee
As a result,  at $T=0$ the quantities in Eqs.\re{2.31}-\re{2.33} can be represented as:  
\begin{eqnarray}
&&\rho _{1}(0)=\dsfrac{1}{2}\sum_{\mathbf{k}}\left(\dsfrac{\veps_{k} +\Delta }{E_{{k}}}-1\right) =
\dsfrac{1}{4\pi^2} \int_{0}^{\infty} {k^2dk} \left(\dsfrac{\veps_{k}
+\Delta }{E_{{k}}}-1\right) \\
\lab{2.36}
&&\sigma (0)=-\dsfrac{\Delta} {4\pi^2} \int_{0}^{\infty}\dsfrac{k^2dk}{E_{k}}\\
\lab{2.37}
&&E_{2}(0)=\dsfrac{1}{4\pi^2} 
\left(\int_{0}^{\infty} E_{{k}}k^{2}dk-\int_{0}^{\infty}\left(\eps_{{k}}+\Delta \right)k^{2}dk\right),
\lab{2.38}
\end{eqnarray}
The divergences in these integrals can be regularized by introducing a cutting
parameter $\Lambda $ ($\Lambda \rightarrow \infty $ at the end
of the calculations) or equivalently by using the dimensional regularization scheme.
Now, assuming for the moment that, for $T=0$ case $\eps_{k}={k}^2/{2m}$
and using dimensional regularization one obtains: 
\footnote{Note that the second integral in Eq. \re{2.38} can be treated with the Veltman formula, see e.g.,
H. Kleinert Hagen and V. Schulte-Frohlinde, {Critical properties of} $\phi^4$-{theories}, World Scientific Publishing Company (2001).}
\begin{eqnarray}
\rho_{1}(0) &=& \dsfrac{(\Delta m)^{3/2}}{3\pi^2},
\lab{2.39}\\
\sigma(0) &=& 3\rho_{1}(0),
\lab{2.40}\\
E_{2}(0) &=& \dsfrac{8(\Delta m)^{5/2}}{15m\pi^2}.
\lab{2.41}
\end{eqnarray}
Summarizing this subsection we rewrite the above formulas as:
\begin{eqnarray}
\rho_{1} &=&\dsfrac{(\Delta m)^{3/2}}{3\pi^2}+\dsint \dsfrac{d^3{k}}{\left( 2\pi \right) ^{3}}f_{B}( E_{{k}})
\dsfrac{\veps_{k}+\Delta }{E_{k}},
\lab{2.42}\\
\sigma &=&\dsfrac{(\Delta m)^{3/2}}{\pi^2}-\Delta \dsint \dsfrac{d^3{k}}{\left( 2\pi \right) ^{3}} f_{B}(
E_{{k}})\dsfrac{1}{E_{{k}}},
\lab{2.43}\\
E&=&\dsfrac{8(\Delta m)^{5/2}}{15m\pi^2}+\dsfrac{U\rho _{0}^{2}}{2}-\mu \rho _{0}-\dsfrac{U}{2}\left( 2\rho _{1}^{2}+\sigma ^{2}\right)+
\dsint \dsfrac{d^3{k}}{\left( 2\pi \right) ^{3}} E_{{k}} f_{B}\left(E_{{k}}\right),
\lab{2.44}
\end{eqnarray}
where Bose distribution of phonons $f_{B}\left( E_{{k}}\right)$ is defined in Eq.\re{2.23}.

To perform the MFA calculations one starts by solving Eqs. \re{2.21} and \re{2.27.2} with
$\rho _{1}$ and $\sigma $ given by Eqs. \re{2.42} and \re{2.43}. As outlined above, in contrast to the BEC
of atomic gases, in the triplon problem the chemical potential $\mu $ is the input
parameter, whereas the densities are the output ones. Bearing this in mind,
we rewrite the main Eqs. \re{2.21} and \re{2.27.2} as
\begin{eqnarray}
\Delta&=&\mu+2U\left(\sigma-\rho_{1}\right),
\lab{2.46}\\
\nonumber
\\
\rho_{0}&=&{\Delta}/{U}-\sigma.
\lab{2.47}
\end{eqnarray}
The system of coupled equations, \re{2.42},
\re{2.43} and \re{2.46}, \re{2.47} has to be solved for given  $T$ and $\mu$ 
to evaluate the triplon density
\be
\rho(T\leq T_{c})=\rho_{0}+\rho_{1}=\dsfrac{\Delta+\mu}{2U},
\lab{2.47.1}
\ee
which is proportional to the measured sample magnetization density $M$. Note that
by formally setting in all above formulas $\sigma\equiv 0$, one arrives at the HFP
approximation and particularly
\begin{equation}
\Delta=\mu -2U\rho_{1}, \qquad \rho_{0}={\Delta}/{U}.
\lab{2.48}
\end{equation}

\subsection{The critical temperature and triplon density}

Before discussing the normal $T>T_{c}$ phase we evaluate  the temperature
of BEC transition. It is well-known that in the MFA the system of interacting Bose condensate
particles at $T\rightarrow T_c$ behaves like an ideal gas. In fact, assuming $\rho_{0}(T\rightarrow T_c)=0$,
$\rho_{1}(T\rightarrow T_c)=\rho_c$,  $\sigma(T\rightarrow T_c)=0$,
$\Delta(T\rightarrow T_c)=0$, and $E_{k}=\veps_k$ one concludes  from Eqs.\re{2.42},\re{2.46} that
\be
\rho_{c}=\dsint\dsfrac{d^3{k}}{(2\pi)^3}
\dsfrac{1}{\exp(\beta_c\veps_{k})-1}=\dsfrac{\mu}{2U}.
\lab{2.49}
\ee
With given $\mu = \mu_{B}gH_{\rm ext}-\Delta_{\rm st}$ and coupling constant $U$,
the critical temperature $T_{c}\equiv 1/\beta_{c}$ can be found as a solution of Eq. \re{2.49}.
Note that for the parabolic dispersion $\veps_{k}=k^{2}/{2m}$,
the integral \re{2.49} can be evaluated analytically giving following well
known relation \ci{pitaevskibook}:
\be
T_{c}^{\rm [par]}=\dsfrac{2\pi}{m}\;\left(\dsfrac{\mu}{2\zeta(3/2)U}\right)^{2/3},
\lab{2.49.1}
\ee
where $\zeta(x)$ is the Riemann function. 

\begin{figure}
\centerline{\includegraphics*[scale=0.6]{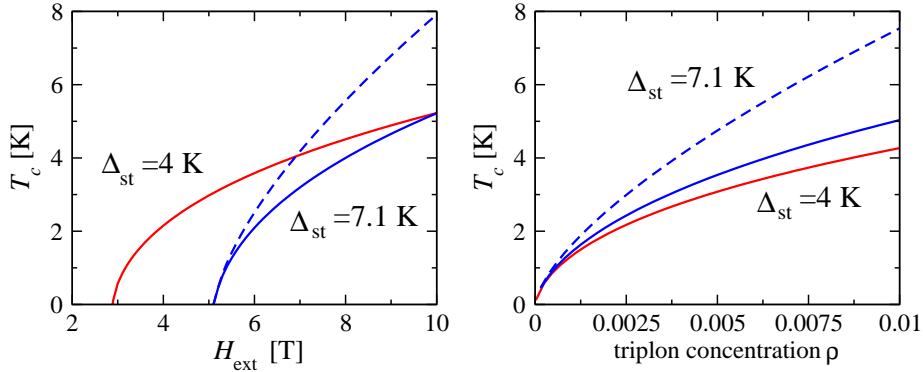}}
\caption{(Color online) The critical temperature of BEC of triplons $T_c$ 
as a function of external field (left panel) and  
corresponding triplon density $\rho=\mu/2U$
(right panel) for two values of the singlet-triplet gap. 
To understand the role of the gap in the magnon spectrum, we present the results
for $\Delta_{\rm st}=4$ K, with the reduced exchange parameter yielding the 
same effective mass as for $\Delta_{\rm st}=7.1$ K. The gap values are marked near the lines. 
The solid and dashed lines are for the relativistic and parabolic dispersions for $\Delta_{\rm st}=7.1$ K,
respectively.}
\label{TCvsH}
\end{figure}

As it has been underlined in Introduction, the bare dispersion of magnons, $\veps_{k}$,
plays the crucial role in determining $T_c$. Figure \ref{TCvsH} presents the transition 
temperature as a function of magnetic field $H_{\rm ext}$ and 
corresponding triplon density $\rho$, that is the sample
magnetization, for parabolic $\veps_{k}={k}^{2}/{2m}$ 
and  "relativistic"  
\be
\veps_{k}=\sqrt{\Delta_{\rm st}^2+J^2k^2/4}-\Delta_{\rm st}
\lab{relat}
\ee
bare dispersion, typical for a system with a gapped spectrum \cite{Kulik11,Grether}.  
Here the effective exchange parameter 
$J=2\sqrt{\Delta_{\rm st}/m}$ is chosen to match the parabolic and the 
relativistic $\varepsilon_{k}$ at small $k$.
The dashed line obtained directly from Eq.\re{2.49.1} displays $T_c\sim\rho^{2/3}$ behavior.
The solid one is a result of numerical solution of Eq.\re{2.49} with the
dispersion in Eq.\re{relat} and  
shows a crossover from  $T_c\sim\rho^{2/3}$ at lower to  
$T_c\sim\rho^{0.5}$ at higher temperatures in agreement with the
experimental data.\footnote{We mention that for linear dispersion $\varepsilon_{k}\sim k$, the
exponent would be equal to $1/3$, and, therefore, in the experimental regime,
the triplon gas is between the nonrelativistic and strongly relativistic 
realizations.}  The decrease in $\Delta_{\rm st}$ enhances the role
of relativistic features in the spectrum and leads to a faster deviation
from the $T_c\sim\rho^{2/3}$ behavior, as can be seen in Fig.\ref{TCvsH}.  
Here and below we mainly use the set of input parameters as  
$m=0.0204$ K$^{-1}$, $\Delta_{\rm st}=7.1$ K, $U=313$ K, and $g=2.06$ 
\ci{Yamada08} valid for the weakly anisotropic quantum antiferromagnet TlCuCl$_{3}$.

In the normal phase the symmetry in Eq.\re{U1} is not broken and the Bogoliubov shift is not needed
either. Here the anomalous density vanishes, $\sigma (T>T_{c})=0$,
and hence, both approximations, HFB and HFP coincide. As a result we obtain for the triplon density
\be
\ba
\rho(T>T_{c})=\dsint \dsfrac{d^3{k}}{\left( 2\pi \right) ^{3}}
\dsfrac{1}{\exp \left( \beta\omega_{k}\right) -1}\\
\lab{2.50}
\ea
\ee
with $\omega_{k}=\veps_{k}-\mu+2U\rho \equiv\veps_{k}-\mu _{\rm eff}$. 
Similarly, by setting 
$\Delta=\rho_{0}=\sigma=0,\ \rho_{1}=\rho,\ E_{{k}}=\omega_{{k}}$ in Eq.\re{2.44}
one obtains following equation for the energy per unit cell
\be
E(T>T_{c})=-U\rho ^{2}+\dsint \dsfrac{d^3{k}}{\left(2\pi\right)^{3}}
\frac{\omega_{k}}{\exp\left(\beta\omega_{k}\right)-1}.
\lab{2.51}
\ee

\begin{figure}
\centerline{\includegraphics*[scale=0.6]{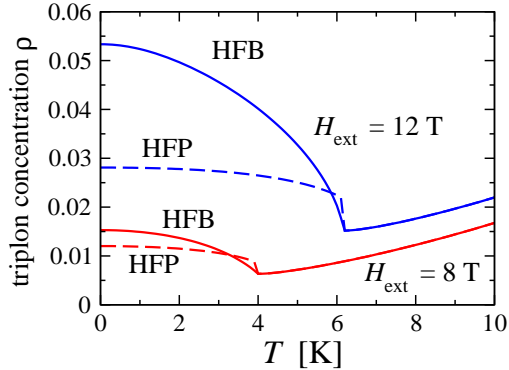}}
\caption{
 (Color online). Comparison of the HFB (solid lines) and the HFP
(dashed lines) results for the triplon density.
The HFB approach shows a continuous behavior, which fully agrees with the experimental
data \cite{Yamada08,Amore08} while the HFP approach leads to the discontinuity.
The corresponding magnetic fields $H_{\rm ext}$ are marked near the plots.}
\label{RHOvsT}
\end{figure}

The density of triplons as a function of temperature, evaluated in the HFP and HFB approximations
is presented in Fig.\ref{RHOvsT}. It is seen that the former leads to a discontinuity in the magnetization
near the critical temperature, while the latter gives a continuous behavior in accordance with the
experimental data \ci{Yamada08,Amore08}. However, in the condensate phase, the triplon density is
higher in the HFB than in the HFP approximation. Therefore, the validity of the description
of the BEC in the HFP approximation can be checked in the magnetization measurement experiments.  

\section{Macroscopic properties: specific heat and magnetic susceptibility}

\subsection{General expressions with nonzero anomalous averages}

Triplon contribution to the constant volume specific heat $C_{v}$,
can be calculated by differentiation of the energy with respect to
the temperature at given chemical potential, that is at given external field \ci{Huangbook,bookLandau}:
\be
C_{v}(H_{\rm ext},T)=\dsfrac{\partial E}{\partial T},
\lab{3.1}
\ee
where the energies for condensate and normal states are given by Eqs. \re{2.44} and \re{2.51},
respectively. 

Before discussing the magnetic susceptibility, we note that
the macroscopic properties of the systems are related to their response 
to external fields. An example is given by the isothermal compressibility
 \be
 \kappa_{T}=-\dsfrac{1}{V}{\left(\dsfrac{\partial V}{\partial P}\right)}_{T}=\dsfrac{1}{\rho}
 {\left(\dsfrac{\partial\rho}{\partial P}\right)}_{T}.
 \lab{1.2}
 \ee
If $\kappa_{T}\rightarrow\infty$ the system becomes unstable \ci{bookLandau,yukalovSTR}
since an infinitesimal fluctuation  of pressure $P$ will lead to its collapse or explosion. 
As to the triplons  their density is proportional
to the magnetization $M$, while the $H_{\rm ext}$ simulates the pressure,
and hence the relevant parameter, which determines the stability 
of the system with respect to the magnetic field, is the susceptibility
\be
\chi(H_{\rm ext},T)=\left(\dsfrac{\partial M}{\partial H_{\rm ext}}\right)_{T}.
\ee
The susceptibility can be calculated with the triplon density as:
\be
\chi(H_{\rm ext},T)=(g\mu_B)^2\dsfrac{\partial\rho}{\partial\mu}.
\lab{3.2}
\ee
We will omit $H_{\rm ext}$ from the arguments of $\chi(H_{\rm ext},T)$ and $C_{v}(H_{\rm ext},T)$ below. 

We begin with the normal phase, where $\rho_0=\sigma=\Delta=0$. The derivative of the density 
with respect to the temperature can be evaluated here with Eq.\re{2.50} as
\be
\frac{\partial \rho}{\partial T}=\dsfrac{\beta S_{1}}{2S_{2}-1}.
\ee
Calculating the derivative of Eq.\re{2.51} with respect to $T$ one obtains:
\be
C_{v}(T>T_{c})=-S_{3}+2US_{1}\frac{\partial\rho}{\partial T}.
\lab{2.52}
\ee
Here we introduced dimensionless quantities
\be
 S_{1}=\sum_{\mathbf{k}}\omega_{k}f_{B}^{\prime}\left( \omega_{k}\right),
 \quad S_{2}=U\sum_{\mathbf{k}}f_{B}^{\prime}\left( \omega_{k}\right),
 \quad S_{3}=\beta\sum_{\mathbf{k}}\omega_{k}^{2}f_{B}^{\prime}\left(\omega_{k}\right),
\ee
and used the relation ${\partial f_{B}(\omega)}/{\partial T}=-\beta \omega f_{B}^{\prime}(\omega)$,
where 
$f_{B}^{\prime}\left(\omega\right)=
{\partial f_{B}(\omega)}/{\partial\omega}=-\beta\exp({\beta\omega})f_{B}^{2}\left(\omega\right)$ 
and $\omega_{k}$ is given in Eq.\re{2.20}. 

Similarly, taking the derivative of the self-consistency Eq.\re{2.50}, which we present here
in the form:
\begin{equation}
\rho (\mu )=\sum_{\mathbf{k}} \dsfrac{1}
{\exp\left[\beta\left(\veps_{k}-\mu +2U\rho(\mu)\right)\right]-1}
\end{equation}
with respect to $\mu$ and solving the equation for $\partial\rho/\partial\mu $ one finds 
$\partial\rho/\partial\mu =S_{2}/U(2S_{2}-1)$. 
As a result, we obtain with Eq.\re{3.2} the normal phase susceptibility:
\begin{equation}
\chi (T>T_{c})=\frac{\left(g\mu_{B}\right)^{2}}{U}
\dsfrac{S_{2}}{2S_{2}-1}.
\lab{2.53} 
\end{equation}

Now we proceed with the condensate phase, where the dependence of $\Delta$ 
and the corresponding normal and anomalous densities on temperature and magnetic field
should be taken into account. 

Here $C_{v}(T)$ is obtained by taking the derivative of 
$E\left(T<T_{c}\right)$ given by Eq. \re{2.44}. The latter can be rewritten as:
\begin{equation}
E\left( T<T_{c}\right) ={E}_{2}-\dsfrac{\mu^{2}}{2U}+U\left( \rho _{1}^{2}-2\rho _{1}\sigma \right),
\end{equation}
where we used the relation $\rho_{0}=\mu/U-2\rho_{1}+\sigma $ and
present ${E}_{2}$ defined in Eq.\re{2.33} as:
\begin{equation}
{E}_{2}=\dsfrac{8m^{3/2}\Delta ^{5/2}}{15\pi^{2}}%
+\sum_{\mathbf{k}}E_{k}f_{B}\left(E_{k}\right).
\end{equation}
As a result:
\begin{equation}
C_{v}(T\leq T_{c})=\dsfrac{\partial E_{2}}{\partial T}-2U\rho _{1}\dsfrac{\partial \sigma }{\partial T}%
-2U\left( \rho -\dsfrac{\mu }{U}\right) \dsfrac{\partial \rho _{1}}{\partial T}.
\lab{2.55}\\
\end{equation}

We begin with calculation of ${\partial\Delta}/{\partial T}$, which is the
key ingredient in the specific heat. It is obtained by differentiating both
sides of the main equation \re{2.46} with respect to the temperature, and
for known ${\partial\Delta}/{\partial T}$ the other two derivatives 
${\partial\rho_{1}}/{\partial T}$ and ${\partial\sigma}/{\partial T}$
can be evaluated directly from Eqs.\re{2.42}, \re{2.43}. As a result
one obtains the derivatives necessary to evaluate the specific heat:
\begin{eqnarray}
&&\dsfrac{\partial \Delta }{\partial T}=
\dsfrac{2U\beta}{1-2U\left(\dsfrac{\sqrt{\Delta}m^{3/2}}{\pi^{2}}-
\sum_{\mathbf{k}}{\cal F}\left(E_{k}\right)\right)}
\sum_{\mathbf{k}}\left(\veps_{k}+2\Delta\right)f_{B}^{\prime}(E_{k}),\\
\lab{2.62}
&&\dsfrac{\partial \rho _{1}}{\partial T}=\dsfrac{m\sqrt{\Delta m}}{2\pi ^{2}}\frac{\partial \Delta}{\partial T}+
\sum_{\mathbf{k}}\left[\dsfrac{\partial\Delta}{\partial T}
\left(\dsfrac{\veps_{k}\Delta}{E_{k}^{2}}{\cal F}\left(E_{k}\right)
+\dsfrac{\veps_{k}^{2}}{E_{k}^{2}}f_{B}^{\prime}\right)-
\beta\left(\veps_{k}+\Delta\right)f_{B}^{\prime}(E_{k})\right],
\lab{2.56.1}
 \\
&&\dsfrac{\partial \sigma }{\partial T}=\dsfrac{3m\sqrt{\Delta m}}{2\pi ^{2}}\frac{\partial \Delta}{\partial T}-
\sum_{\mathbf{k}}
\left[\frac{\partial\Delta}{\partial T}
\left(
\dsfrac{\veps_{k}\Delta}{E_{k}^{2}}{\cal F}\left(E_{k}\right)
+\dsfrac{\veps_{k}^{2}}{E_{k}^{3}}f_{B}(E_{k})
\right) 
-\beta \Delta f_{B}^{\prime}(E_{k})\right],
\lab{2.56.2}
\end{eqnarray}
where we introduced notation for the frequently used expression:
\begin{equation}
{\cal F}\left(E_{k}\right)\equiv\dsfrac{f_{B}(E_{k})}{E_{k}}+f_{B}^{\prime}(E_{k}).
\lab{FE}
\end{equation}
The expression for 
\begin{equation}
\dsfrac{\partial E_{2}}{\partial T}=\dsfrac{4\left( \Delta m\right) ^{3/2}}{3\pi ^{2}}
\dsfrac{\partial \Delta }{\partial T}+\sum_{\mathbf{k}}\left(\frac{\partial\Delta}{\partial T}
\veps_{k}{\cal F}\left(E_{k}\right) -\beta E_{k}^{2}f_{B}^{\prime}(E_{k})\right)
\lab{2.56.3}
\end{equation}
completes the set of derivatives necessary for evaluation of the specific heat. 

The derivative ${\partial\Delta}/{\partial\mu}$ can be found by  differentiating both sides of the main
equation \re{2.46} with respect to $\mu$.
As a result,
\begin{equation}
\dsfrac{\partial \Delta }{\partial \mu }=\dsfrac{1}
{1-2U\left(\dsfrac{\sqrt{\Delta}m^{3/2}}{\pi^{2}}-\sum_{\mathbf{k}}{\cal F}\left(E_{k}\right)
\right)}.
\lab{2.57}
\end{equation}
As to ${\partial\rho}/{\partial\mu}$, needed for evaluation of the magnetic susceptibility
in Eq.\re{3.2}, it is obtained by
differentiating the equation $\rho =\left(\Delta+\mu \right) /2U,$
with respect to $\mu$. This yields
\be
\chi(T\leq T_{c})=\dsfrac{\left(g\mu_{B}\right)^{2}}{2U}
\left(\dsfrac{\partial \Delta }{\partial \mu }+1\right).
\lab{2.56}
\ee
Below we apply these equations to the cusps in the specific heat
and susceptibility and to the qualitative effects such as the instability 
in strong magnetic fields.  We will show that at a given temperature, 
there exists a critical field $H^{\rm cr}_{\rm ext}(T)$, such that when 
$H_{\rm ext}$ approaches $\geq H^{\rm cr}_{\rm ext}(T)$,
the magnetic susceptibility, $\chi(H_{\rm ext},T)$ diverges.

\subsection{Cusp in the specific heat and magnetic susceptibility near $T_{c}$}

The cusp in the specific heat defined as
\begin{equation}
\Delta C_{v}=\lim_{T\rightarrow T_{c}-0}C_{v}(T)-\lim_{T\rightarrow
T_{c}+0}C_{v}(T)
\lab{3.17}
\end{equation}
is an interesting quantity in the theory of phase transitions. In accordance with 
the Ehrenfest classification, a phase transition with the discontinuity
in $C_{v}$ near the transition point, is the second order one.
Particularly, it is well-known that \ci{Huangbook,bookLandau} for the ideal gas $\Delta C_{v}$
near the transition into BEC is zero i.e.  the specific heat is continuous. 
We shall illustrate this fact for completeness.
The specific heat per particle of the ideal Bose gas is given by \ci{Huangbook,bookLandau}
\be
\ba
{C_{v}^{U=0}(T\leq T_c)}=\dsfrac{15\zeta(5/2)}{4\zeta(3/2)}
\left(\dsfrac{T}{T_{c}}\right)^{3/2},\\
\lab{3.18}
\ea
\ee
\be
\ba
{C_{v}^{U=0}(T> T_c)}=\dsfrac{15g_{5/2}(z)}{4g_{3/2}(z)}-
\dsfrac{9g_{3/2}(z)}{4g_{1/2}(z)},
 \lab{3.19}
\ea
\ee
where $g_{p}(z)$ is defined as:
\begin{equation}
g_{p}(z)=\dsfrac{1}{\Gamma \left( p\right) }\int_{0}^{\infty }dx\dsfrac{x^{p-1}%
}{z^{-1}e^{x}-1},
\lab{3.20}
\end{equation}
related to the Riemann function as $\zeta(p)=g_{p}(z=1)$. 
When $T\rightarrow T_{c}+0$, the fugacity $z=\exp({\beta\mu})<1$ tends to unity,
i. e. $\ds \lim_{T\rightarrow T_{c}+0}z=1$ and the second term in \re{3.19} vanishes,
because of the divergence in $g_{1/2}$, i.e.
$g_{1/2}(z)\sim(1-z)^{-1/2}$ in this region, while
the first term exactly coincides with \re{3.18}. As a result,
\be
\Delta C_{v}^{U=0}=0,
\lab{3.21}
\ee
and hence the specific heat of the ideal gas is continuous although being plotted as a 
function of temperature $C_{v}^{U=0}(T)$ behaves similarly  to the $\lambda$ curve.

We proceed with calculation of $\Delta C_{v}$.
Bearing in mind that for $T>T_{c}$ HFB and HFP approximations coincide, from \re{2.52} we obtain
\begin{eqnarray}
\lim_{T\rightarrow T_{c}+0}C_{v}(T)=-\beta_{c} \sum_{\mathbf{k}}\veps_{k}^{2}f_{B}^{\prime}(\veps_{k}).
\lab{3.27}
\end{eqnarray}

For $T<T_{c},$ $C_{v}$ is given by Eqs. \re{2.55}-\re{FE}. Assuming in the last equations  
$E_{{k}}=\veps_{k}$ and $ \Delta=0,$ one finds
\begin{equation}
\lim_{T\rightarrow T_{c}-0}\frac{\partial \Delta}{\partial T}=
\dsfrac{2\beta_{c}U}{1+2U\ds{\sum_\mathbf{k}}
{\cal F}\left(\varepsilon_{k}\right)}
{\sum_\mathbf{k}}f_{B}^{\prime}(\veps_{k})\veps_{k},
\lab{3.28}
\end{equation}
which is finite.
Using \re{3.28} in \re{2.56.1}-\re{FE} and setting in \re{2.55} $\rho=\mu/2U$, $\rho_{0}=0,\rho_{1}=\rho,\sigma=0$, 
one obtains for the maximum value of the specific heat $\mbox{max}\{C_{v}^{\rm HFB}(T)\}\equiv\lim_{T\rightarrow T_{c}-0}C_{v}(T)$: 
\begin{equation}
\lim_{T\rightarrow T_{c}-0}C_{v}(T)=
\sum_{\mathbf{k}}\left( \mu +\veps_{k}\right) 
\left(
{\cal F}\left(\varepsilon_{k}\right)
\frac{\partial \Delta}{\partial T}-
\beta_{c} \veps_{k}f_{B}^{\prime}(\veps_{k})\right). 
\lab{3.29}
\end{equation}
Subtracting \re{3.27} from \re{3.29} we finally obtain
\begin{equation}
\Delta C_{v}=\sum_{\mathbf{k}}
\left(
{\cal F}\left(\varepsilon_{k}\right) 
\left(\mu+\veps_{k}\right)
\frac{\partial \Delta}{\partial T} -\mu \beta_{c} \veps_{k}f_{B}^{\prime}(\veps_{k})
\right),
\lab{3.30}
\end{equation}
where ${\partial\Delta}/{\partial T}$ is given by Eq. \re{3.28}.

In a quite similar way one can calculate the cusp in the susceptibility:
\begin{equation}
\Delta\chi=\dsfrac{\left({g}\mu_{B}\right)^{2}}{2U}
\frac{1}{1+2U\displaystyle{\sum_{\mathbf{k}}}{\cal F}\left(\varepsilon_{k}\right)}.
\lab{3.31}
\end{equation}

\subsection{Parabolic dispersion, $\veps_{k}={k}^{2}/2m$}

For the parabolic dispersion, calculations can be done analytically, as presented below.
The specific heat is expressed as:
\begin{eqnarray}
&&\mbox{max}\{C_{v}(T)\}  =-\beta_{c} R_{1}-\mu\beta_{c} R_{2}+
\frac{\partial\Delta}{\partial T}\left(R_{2}+\mu R_{3}+R_{5}\right),
\lab{3.34}\\
&&\lim_{T\rightarrow T_{c}+0}C_{v}(T) =-\beta_{c} R_{1},
\lab{3.35} \\
&&\Delta C_{v} =\mbox{max}\{C_{v}(T)\} +\beta_{c} R_{1}=\frac{\partial \Delta}{\partial T}
\left(R_{2}+\mu R_{3}+R_{5}\right)-\mu \beta_{c} R_{2},
\lab{3.36}
\end{eqnarray}
and the cusp in the susceptibility has the form:
\begin{equation}
\Delta \chi  =\frac{\left(g\mu_{B}\right)^{2}}{2U}\frac{\partial \Delta}{\partial \mu}.
\lab{3.38}
\end{equation}
The derivatives ${\partial \Delta}/{\partial T}$ and ${\partial\Delta}/{\partial\mu}$
can be simplified to:
\begin{eqnarray}
\frac{\partial \Delta}{\partial T} &=&\dsfrac{2\beta_{c}U R_{2}}{1+2UR_{3}},
\lab{3.39} \\
\frac{\partial \Delta}{\partial \mu}&=&\dsfrac{1}{1+2UR_{3}}.
\lab{3.40}
\end{eqnarray}
In these formulas we used following notations:
\begin{eqnarray}
R_{1} &=&-\beta_{c} \sum_{\mathbf{k}}f_{B}^{2}(\veps_{k})\veps_{k}^{2}e^{\beta_{c}
\veps_{k}}=-\dsfrac{\zeta(5/2) \Gamma \left( 7/2\right) }{\sqrt{2}\pi ^{2}}
T_{c}k_{T}^{3},
\lab{3.41} \\
R_{2} &=&-\beta_{c} \sum_{\mathbf{k}}f_{B}^{2}(\veps_{k})\veps_{k}e^{\beta_{c}
\veps_{k}}=-\dsfrac{\zeta(3/2)
 \Gamma \left( 5/2\right) }{\sqrt{2}\pi ^{2}}k_{T}^{3},
 \lab{3.42} \\
R_{3} &=&\sum_{\mathbf{k}}\left( f_{B}^{\prime}(\veps_{k})+f_{B}(\veps_{k})/\varepsilon_{k}\right)=-\dsfrac{0.6471\sqrt{2}}{\pi ^{2}}
\frac{k_{T}^{3}}{T_{c}},
\lab{3.43} \\
R_{4} &=&\sum_{\mathbf{k}}f_{B}(\veps_{k})\veps_{k}=
\dsfrac{\zeta(5/2)\Gamma\left(5/2\right)}{\sqrt{2}\pi^{2}}T_{c}k_{T}^{3},
\lab{3.44} \\
R_{5} &=&\sum_{\mathbf{k}}f_{B}(\veps_{k})=\dsfrac{\zeta(3/2)\Gamma\left(3/2\right)}{\sqrt{2}\pi^{2}}k_{T}^{3},
\lab{3.45}
\end{eqnarray}
where we introduced $k_{T}\equiv\sqrt{mT_{c}}$ for the characteristic thermal wavevector of
triplon at the transition temperature. 
Taking into account that $R_{5}$ is the triplon concentration, we find that 
Eq.\re{3.45} is equivalent to Eq.\re{2.49.1}.

\section{High- field instability}

It is well-known that \ci{yukalovSTR,ueda} the dynamic stability of equilibrium system
is determined by its excitation spectrum: When the latter becomes imaginary
the time evolution of excitations changes qualitatively. 
In the condensate state the spectrum is given by  $E_{k}=\sqrt{\veps_{k}}\sqrt{\veps_{k}+2\Delta}$
where $\Delta$ is the solution of nonlinear algebraic Eq.\re{2.46} which may be rewritten
as follows
\be
\Delta=\mu+\dsfrac{4U(\Delta m)^{3/2}}{3\pi^2}-\frac{U}{\pi^{2}}\int_{0}^{\infty}
\dsfrac{\veps_{k}+2\Delta}{E_{k}}f_B(E_{k})k^{2}dk.
\lab{5.1}
\ee
Obviously, for some set of system parameters $U$, $m$ 
and external fields $H_{\rm ext}$, Eq.\re{5.1} may have no real solution
and hence the  spectrum, as well as the sound speed,  become imaginary.
For example, in our previous work we have shown that for parameters \ci{Yamada08} this may happen,  at $H_{\rm ext}^{\rm cr}=12.5$ T at $T=0$ with the increasing in $H_{\rm ext}^{\rm cr}$ with $T$. 
Below we investigate the magnetic susceptibility near the instability point and show that it goes to infinity, as expected
in these cases, providing the direct experimental test for the instability. 
This effect can be seen from the fact that when $H_{\rm ext}=H_{\rm ext}^{\rm cr}$, i. e. $\mu=\mu_{\rm cr}$
the  denominator of Eq. \re{2.57} becomes zero, and hence $\partial\Delta/\partial\mu$ diverges. 
For $\mu>\mu_{\rm cr}$ the main equation \re{5.1} has no  positive solution and hence the sound speed $c=\sqrt{\Delta}/\sqrt{m}$ becomes complex. 
Below we illustrate this analytically for $T=0$.

\begin{figure}
\centerline{\includegraphics*[scale=0.6]{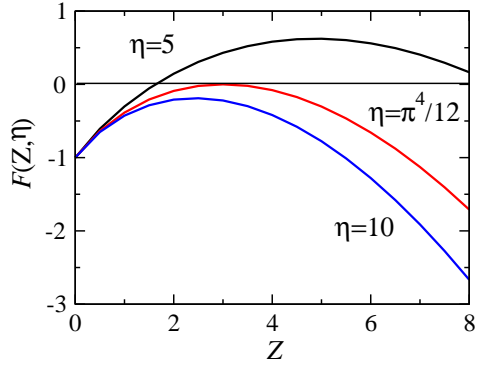}}
\caption{The LHS of Eq. \re{5.5} for different values of $\eta$. The real  solutions of
 this equation correspond to the intersection of $F(Z,\eta)$ with the $Z=0$ axis.}
\label{LHSvsZ}
\end{figure}

For zero temperature the derivative ${\partial\Delta}/{\partial\mu}$ in Eq. \re{2.57} is written as
\be
\frac{\partial\Delta}{\partial\mu}=\dsfrac{1}{1-\dsfrac{2U\sqrt{\Delta}m^{3/2}}{\pi^{2}}},
\lab{5.2}
\ee
where $\Delta$ is the solution to the following equation:
\be
\Delta=\mu+\dsfrac{4U(\Delta m)^{3/2}}{3\pi^{2}}.
\lab{5.3}
\ee
Introducing dimensionless variables $\eta = \mu U^{2} m^{3}$ and $Z=\Delta /\mu$ one can rewrite
the last two equations  as:
\bea
&& \frac{\partial\Delta}{\partial\mu}=\dsfrac{1}{1-\dsfrac{2\sqrt{\eta Z}}{\pi^{2}}},
\lab{5.4}\\
&& F(Z,\eta)\equiv Z-\dsfrac{4Z^{3/2}\sqrt{\eta}}{3\pi^{2}}-1=0,
\lab{5.5}
\eea
where $\eta$ is an input parameter.
In Fig.\ref{LHSvsZ} $F(Z,\eta)$ is plotted as a function of $Z$ for various values of $\eta$.
It is seen that for $\eta$ larger than some  critical $\eta_{\rm cr}$,
the  second term in \re{5.5} dominates, $F(Z,\eta)$ is always negative, and hence there is no real solutions.
If the maximum value of $F(Z,\eta)$, $(\max{F(Z,\eta)})$,
for a given $\eta$ is negative the Eq. \re{5.5}, has no real solution.
Otherwise, when $\max(F(Z,\eta)$ is positive, $F(Z,\eta)$ intersects the $Z$-axis and
the equation has at least one positive solution.
Thus, the boundaries $\eta_{\rm cr}$ and $Z_{\rm cr}$ are defined by the coupled equations:
\begin{eqnarray}
&&F(Z_{\rm cr},\eta_{\rm cr})= Z_{\rm cr}-\dsfrac{4Z_{\rm cr}^{3/2}\sqrt{\eta_{\rm cr}}}{3\pi^{2}}-1=0,
\lab{5.6}\\
&&\left.\frac{\partial F}{\partial Z}\right|_{Z=Z_{\rm cr},\eta=\eta_{\rm cr}}
=1-\dsfrac{2\sqrt{Z_{\rm cr}\eta_{\rm cr}}}{\pi^{2}}=0,
\lab{5.7}
\end{eqnarray}
giving $\eta_{\rm cr} = \pi^{4}/12=8.1174$ and $Z_{\rm cr}=3$.
By comparing \re{5.7} and \re{5.4} one concludes that when 
$\eta$ approaches $\eta_{\rm cr}$, ${\partial\Delta}/{\partial\mu}$ goes to infinity and so  does the
magnetic susceptibility given by Eq. \re{2.56}  i.e. $\chi$ can diverge.
When $\eta$ exceeds $\eta_{\rm cr}$, solutions of Eqs. \re{5.3}, \re{5.5} acquire an imaginary part.
Bearing in mind that $\eta=\mu m^{3}U^{2}=(\mu_{B}gH_{\rm ext}-\Delta_{\rm st})m^{3}U^{2}$, one concludes 
that even at $T=0$, if the $H_{\rm ext}$ is strong enough the speed of sound $c=\sqrt{\Delta/m}=\sqrt{\mu Z/m}$ 
becomes complex and the Bose condensed system of triplons displays dynamical instability.
By expanding Eq.\re{5.5} in the vicinity of the point $Z=Z_{\rm cr}$,$\eta=\eta_{\rm cr}$,
we obtain in this region for the real and imaginary parts of the speed of sound:
${\rm Re}(c)=\sqrt{3\mu_{\rm cr}/m}$ and ${\rm Im}(c)/{\rm Re}(c)=-2\sqrt{\mu-\mu_{\rm cr}}Um^{3/2}/\pi^{2}$,
where $\mu_{\rm cr}\equiv\eta_{\rm cr}/(m^{3}U^{2})$.

\begin{figure}
\centerline{\includegraphics*[scale=0.6]{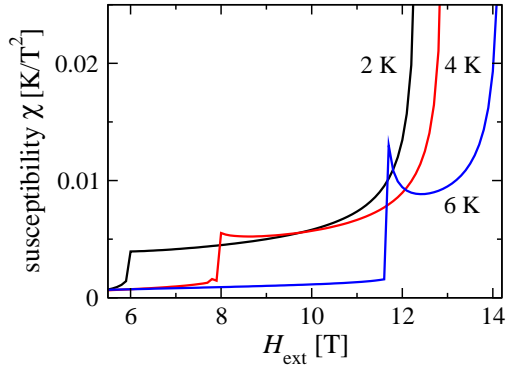}}
\caption{The magnetic susceptibility $\chi(T,H_{\rm ext})$ for the temperatures marked near
      the plots. When external magnetic field reaches the critical value $H_{\rm ext}^{\rm cr}$
      the susceptibility diverges.}
\label{CHIvsHext}
\end{figure}

\begin{figure}
\centerline{\includegraphics*[scale=0.6]{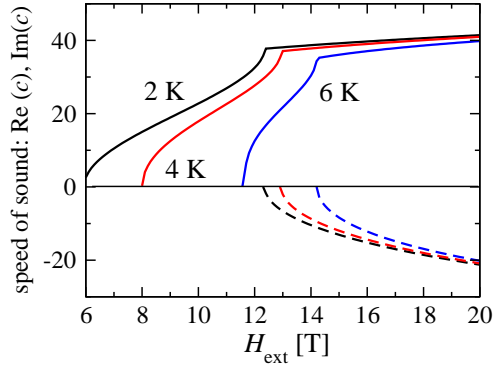}}
\caption{The components of the sound speed $c$ for different temperatures, real for
      $H_{\rm ext}<H_{\rm ext}^{\rm cr}$, and complex out of this range. Real parts ${\rm Re}(c)$ (solid lines) are positive
and ${\rm Im}(c)$ (dashed lines) are negative. Near the threshold
${\rm Im}(c)$ demonstrates to the mean-field square root behavior as discussed 
below Eq.\re{5.7}.}
\label{SOUNDvsHext}
\end{figure}

Similarly it can be shown that for any given $T\leq T_{c}$ there is a maximal value of the external magnetic 
field $H_{\rm ext}^{\rm cr}$ such that the system posses dynamical instability at this point i.e.
$\chi(H_{\rm ext}^{\rm cr},T^{\rm cr})\rightarrow\infty$ and simultaneously the sound speed acquires
an imaginary part.
This fact is illustrated in Figs.\ref{CHIvsHext} and \ref{SOUNDvsHext}. In Fig.\ref{CHIvsHext} the susceptibility 
$\chi(H_{\rm ext},T)$ is plotted as a function of $H_{\rm ext}$. 
In Fig.\ref{SOUNDvsHext} the solution of the main equation, more precisely, the 
components of the sound speed $c$ are plotted  as a function of $H_{\rm ext}$ 
for several temperatures. It is clearly seen that for a given 
temperature the $\chi(H_{\rm ext},T)$ diverges and the
sound speed becomes complex at the same $H_{\rm ext}$.

Note that in the HFP approximation Eq.\re{5.5} being  written as
\be
F_{\rm HFP}(Z,\eta)=Z+\dsfrac{2Z^{3/2}\sqrt{\eta}}{3\pi^{2}}-1=0
\lab{5.8}
\ee
has a positive solution for any positive $\eta$. Thus this approximation precludes the 
appearance of the instability.

The experiment related results from the observed instability are presented in Fig.\ref{CHIvsHext}
and Fig.\ref{SOUNDvsHext}.  Figure \ref{CHIvsHext} shows the divergence of the susceptibility.
Figure \ref{SOUNDvsHext} shows how the imaginary part in the velocity appears and the real part 
of the velocity changes if the field becomes stronger than the threshold value. This is a qualitative
effect, which should demonstrate itself in the experiment. Here an important comment on the
values of the fields where the singular behavior of $\Delta$ can be observed is in order. A
comparison of the calculations presented in Fig.\ref{RHOvsT}, experimental results \cite{Yamada08,Amore08},
and theory taking into account the exact spectrum of triplons \cite{Misguich04}, shows that our model
spectrum leads to an overestimate of the tiplon concentration by about a factor of 1.5. 
Therefore, it underestimates the threshold field, leading to the result that fields of the 
order of 20 T will be necessary \cite{Tanaka01} to put the condensate in the unstable part 
of the phase diagram. 

We mention two characteristic features of the instability considered in this Section.
First, it is not a hydrodynamical instability arising in the inhomogeneous 
flow of the Bose-Einstein 
condensate \cite{Creffield,Modugno,Sinha,Argaman}. Second, it is not directly related to 
the phonon-phonon scattering \cite{Chung} since
we are still in the mean field approximation, and the Hamiltonian we consider
does not contain higher-order products of the phonon operators.  Nevertheless,
both these effects can be studied taking into account the anomalous
averages in the theory. 

\begin{figure}
\centerline{\includegraphics*[scale=0.6]{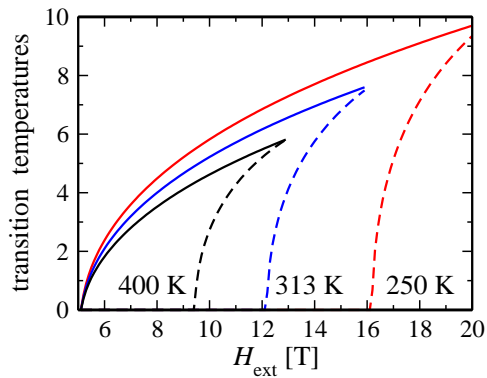}}
\caption{The instability borders (dashed lines) as well as the condensate formation
borders (solid lines) for different $U$. With the increase in $U$ the area corresponding
to the stable condensate decreases.}
\label{phasediagram}
\end{figure}

Summarizing this Section we conclude that there are two characteristic 
values of $H_{\rm ext}$.
First, when the triplons form the BEC,  $H_{\rm ext}^{\rm BEC}$ and
second, $H_{\rm ext}^{\rm cr}>H_{\rm ext}^{\rm BEC}$, when the BEC displays 
dynamical instability. 
The phase diagrams for different repulsion parameters $U$ for relativistic
triplon dispersion are shown in Fig.\ref{phasediagram}. 
The localization of the stable BEC phases on the $(T,H_{\rm ext})$ plane depends
on the model parameters $(m,U,\Delta_{\rm st})$. One can conclude that increasing
of $U$ decreases $H_{\rm ext}^{\rm BEC}$ and makes the BEC less stable, as expected 
from the fact that the instability is caused by the magnon-magnon repulsion.

 \section{Conclusions}

We investigated macroscopic properties of the triplon condensates
taking into account the anomalous density averages.  We show that these
averages have an important quantitative effect on the  specific heat and
qualitatively modify the magnetic susceptibility of these systems. 
The qualitative result of our consideration is related to
the dynamical instabilities arising in the condensate. These instabilities
are seen as the divergence in the magnetic susceptibility and changes
in the sound of the Bogoliubov mode when the external magnetic field becomes
stronger than the corresponding temperature-dependent critical value. 
In this domain, the speed of sound acquires an imaginary part and dependence of its 
real part on the external field becomes weak, almost flat. These two predictions can easily
be verified experimentally in laboratory studies of the static magnetic susceptibility
and in the neutron scattering measurements in various magnetic fields.  

To conclude this paper, one general comment should be given. 
Several authors \cite{Mills,Loktev} casted doubts on the applicability of the 
Bose-Einstein condensation approach to magnetic transitions. The irrefutable proof
of the validity of this approach can be given by the observation of the Josephson
effect, as it was done for magnons in the superfluid He$^3$ \cite{Volovik} and although recently 
proposed for magnetic insulators \cite{Schilling}, not yet observed for the triplon realization. 
However, we believe that observation of the effects predicted in this paper 
which are solely based on the improved theory
of the Bose-Einstein condensate, will resolve this controversy.

{\bf Acknowledgement.} AR and SM acknowledge support of the Volkswagen Foundation.
This work of EYS was supported by the University of Basque Country UPV/EHU grant GIU07/40,
MCI of Spain grant FIS2009-12773-C02-01, and "Grupos Consolidados UPV/EHU 
del Gobierno Vasco" grant IT-472-10. We are grateful to H. Kleinert, A. Pelster, M. Modugno,  
and O. Tchernyshyov for valuable discussions.

\bb{99}

\bi{Matsuda} T. Matsubara and H. Matsuda, Prog. Theor. Phys. {\bf 16}, 569 (1956); 
            H. Matsuda and T. Tsuneto, ibid {\bf 46}, 411 (1970). 

\bi{Tachiki}  M. Tachiki and T. Yamada, 
              J. Phys. Soc. Jpn. {\bf 28}, 1413 (1970).

\bi{Batyev} E.G. Batyev and S.L. Braginskii, JETP {\bf 60}, 781 (1984); 
            E.G. Batyev, {\it ibid} {\bf 62}, 173 (1985). 

\bi{Demidov08} For the BEC in magnetic systems under external pumping:
               V. E. Demidov O. Dzyapko, S. O. Demokritov, G. A. Melkov, and A. N. Slavin,
               Phys. Rev. Lett. {\bf 100}, 047205 (2008).
               For theory:
               Yu. D. Kalafati and V. L. Safonov, JETP Lett. {\bf 50}, 149 (1989); 
               I. S. Tupitsyn, P. C. Stamp, and A. L. Burin,
               Phys. Rev. Lett. {\bf 100}, 257202 (2008);
               A. I. Bugrij and V. M. Loktev, Low Temp. Phys. {\bf 33}, 37 (2007).

\bi{Volovik}   G.E. Volovik,
               J. Low. Temp. Phys. {\bf 153} 266 (2008). 

\bi{Affleck} I. Affleck, Phys.Rev B {\bf 43}, 3215 (1991).

\bi{Giamarchi99} T. Giamarchi and A. M. Tsvelik,
		 Phys. Rev. B {\bf 59}, 11398 (1999).

\bi{Giamarchi08} T. Giamarchi, C. R\"{u}egg, and  O. Tchernyshyov,
                 Nature Physics {\bf 4}, 198 (2008).

\bi{Nikuni00}  T. Nikuni, M. Oshikawa, A. Oosawa, and H. Tanaka, 
               Phys. Rev. Lett. {\bf 84}, 5868 (2000).

\bi{oosawa} A. Oosawa, H. A. Katori, and H. Tanaka, Phys. Rev. B {\bf 63}, 134416, (2001).

\bi{Ruegg07} Ch. C. R\"{u}egg, D. F. McMorrow, B. Normand, H. M. Ronnow, S. E. Sebastian, I. R. Fisher, 
              C. D. Batista, S. N. Gvasaliya, Ch. Niedermayer, and J. Stahn,  
              Phys. Rev. Lett. {\bf 98}, 017202 (2007). 

\bi{Stone07} M. B. Stone, C. Broholm, D. H. Reich, P. Schiffer, O. Tchernyshyov, P. Vorderwisch, and N. Harrison,
             New J. Phys. {\bf 9},  31 (2007).	

\bi{Aczel09} A. A. Aczel, Y. Kohama, M. Jaime, K. Ninios, H. B. Chan, 
             L. Balicas, H. A. Dabkowska, and G. M. Luke,
	     Phys. Rev. B {\bf 79}, 100409(R) (2009).

\bi{Paduan09}  A. Paduan-Filho, K. A. Al-Hassanieh, P. Sengupta, and M. Jaime,
	       Phys. Rev. Lett. {\bf 102}, 077204 (2009).
	
\bi{Laflorencie09} N. Laflorencie and F. Mila,
                   Phys. Rev. Lett. {\bf 102}, 060602 (2009).

\bi{Tsirlin} A. A. Tsirlin and H. Rosner, Phys. Rev. B {\bf 83}, 064415 (2011). 

\bi{Kim} T. Dodds, B.-J. Yang, and Y.B. Kim,
         Phys. Rev. B  {\bf 81}, 054412 (2010).

\bi{Sirker04}   J. Sirker, A. Wei{\ss}e, and O. P. Sushkov,
                 Europhys. Lett. {\bf 68}, 275 (2004).

\bi{Stanescu08}  T. D. Stanescu, B. Anderson, and V. Galitski,
                 Phys. Rev. A {\bf 78}, 023616 (2008).

\bi{Continentino} M.A. Continentino and A.S. Ferreira,
                 Journal of Magn. and Magn. Mat. 
                 {\bf 310}, 828 (2007).

\bi{Sherman03} E. Ya. Sherman, P. Lemmens, B. Busse, A. Oosawa, and H. Tanaka,
               Phys. Rev. Lett. {\bf 91}, 057201 (2003)

\bi{Oosawa04}  A. Oosawa, K. Kakurai, T. Osakabe, M. Nakamura, 
               M. Takeda and H. Tanaka,
               J. Phys. Soc. Jpn. {\bf 73} 1446 (2004).

\bi{Kulik11} Y. Kulik and O. P. Sushkov, arXiv:1104.1245 (unpublished).

\bi{Ruegg} Ch. R\"{u}egg, N. Cavadini, A. Furrer, H.-U. G\"{u}del, K. Kr\"{a}mer, 
           H. Mutka, A. Wildes, K. Habicht, and P. Vorderwisch, 
           Nature {\bf 423}, {\bf 62} (2003).

\bi{ourprb} A. Rakhimov, E. Ya. Sherman and Chul Koo Kim Phys. Rev. B {\bf 81}, 020407(R) (2010).

\bi{Crisan05}  M. Crisan I. Tifrea, D. Bodea, and I. Grosu,
               Phys. Rev. B {\bf 72}, 184414 (2005) provided renormalization group 
               analysis of the triplons BEC. 

\bi{yukANN} V. I. Yukalov, Ann. Phys. {\bf 323}, 461 (2008)

\bi{Misguich04} G. Misguich and M. Oshikawa,
                J. Phys. Soc. Jpn. {\bf 73}, 3429 (2004)

\bi{Jensen} J. Jensen, Phys. Rev. B {\bf 83}, 064420 (2011). 

\bi{Keeling} Similar important dispersion-related effects are expected for the Bose-condensation 
             of polaritons: J. Keeling, Phys. Rev. B {\bf 74}, 155325 (2006).

\bi{Yamada08} F. Yamada, T. Ono, H. Tanaka, G. Misguich, M. Oshikawa, and T. Sakakibara,
              J. Phys. Soc. Jpn. {\bf 77}  013701 (2008).

\bi{prokjack}    N.P. Proukakis and B. Jackson, J. Phys. B: At. Mol. Opt. Phys. {\bf 41}, 203002 (2008).

\bi{Kleinert05}    H. Kleinert, S. Schmidt, and A. Pelster, 
                   Annalen der Physik (Leipzig), {\bf 14}, 214 (2005).

\bi{AndersenRevModPhys} Jens O. Andersen,
                     Rev. Mod. Phys. {\bf 76}, 599 (2004)

\bi{yukanom} V.I. Yukalov and E.P. Yukalova, Laser Phys. Lett. {\bf 2}, 506 (2005).

\bi{Yukalov11} V. I. Yukalov, A. Rakhimov, and S. Mardonov,  Laser Physics {\bf 21}, 264 (2011). 

\bi{stoofbook} H. T. C. Stoof, K. B. Gubbels and D.B.M. Dickerscheid
{\it Ultracold Quantum Fields} (Springer, 2009)

\bi{MYPRA} A. Rakhimov, Chul-Koo Kim, Sang-Hoon Kim, and Jae-Hyung Yee, Phys. Rev. A {\bf 77}, 033626 (2008)

\bi{Dickhoffbook} W. H. Dickhoff and D. Van Neck,
                 {\it Many-Body Theory Exposed}
	          World Scientific (2005).

\bi{booknikinu}  A. Griffin,  T. Nikuni, and E. Zaremba,
                {\it Bose-Condensed Gases at Finite Temperatures}
                 Cambridge University Press (2009).

\bi{pitaevskibook} L. Pitaevskii and S. Stringari,  
                  {\it Bose-Einstein Condensation (International Series of Monographs on Physics)}
                  Oxford University Press (2003).

\bi{Grether} Theory of BEC in ideal relativistic gas was developed 
             in M. Grether, M. de Llano, and G. A. Baker, Jr.,  
             Phys. Rev. Lett. {\bf 99}, 200406 (2007)  

\bi{Amore08} R. Dell'Amore, A. Schilling, and K. Kr\"{a}mer
             Phys. Rev. B {\bf 79}, 014438 (2009);
             R. Dell'Amore, A. Schilling, and K. Kr\"{a}mer, 
             Phys. Rev. B {\bf 78}, 224403 (2008).

\bi{Huangbook}   Kerson Huang, {\it Statistical Mechanics} Wiley (1987).

\bi{bookLandau} L.D. Landau and E.M. Lifshitz,  {\it Statistical Physics}, 
                Course of Theoretical Physics, Volume 5, Butterworth-Heinemann Publ. (1980).

\bi{yukalovSTR} V. I. Yukalov, Phys. Rev. E {\bf 72}, 066119 (2005) and references therein. 

\bi{ueda} D. C. Roberts and M. Ueda, Phys. Rev. A {\bf 73}, 053611 (2006) perfomed 
           stability analysis for a multicomponent BEC in terms different from those presented
           here.

\bi{Tanaka01} {For example, it would be of interest to obtain the neutron scattering measurements data 
              H. Tanaka, A. Oosawa, T. Kato, H. Uekusa, Y. Ohashi, K. Kakurai, and A. Hoser, 
             J. Phys. Soc. Jpn. {\bf 70}, 939 (2001) for the fields $H_{\rm ext}$ higher than 15 T.}

\bi{Creffield} C.E. Creffield, 
               Phys. Rev. A {\bf 79}, 063612 (2009). 

\bi{Modugno} M. Modugno, C. Tozzo, and F. Dalfovo,
             Phys. Rev. A {\bf 70}, 043625 (2004).  

\bi{Sinha} S. Sinha and Y. Castin,
           Phys. Rev. Lett. {\bf 87}, 190402 (2001). 

\bi{Argaman} N.  Argaman and Y.B. Band, Phys. Rev. A {\bf 83}, 023612 (2011), included
             anomalous density terms in the density functional theory
             approach. 

\bi{Chung} Ming-Chiang Chung and A. B. Bhattacherjee,
           New J. Phys. {\bf 11}, 123012 (2009).

\bi{Mills} D. L. Mills,  Phys. Rev. Lett. {\bf 98}, 039701 (2007).

\bi{Loktev}  V. M. Kalita, I. Ivanova, and V. M. Loktev, Phys. Rev. B {\bf 78}, 104415 (2008),
V. M. Kalita and V. M. Loktev,  JETP Letters, {\bf 91}, 183 (2010), 
V. M. Kalita and V. M. Loktev, Low Temp. Phys. {\bf 36}, 665 (2010).

\bi{Schilling} A. Schilling and H. Grundmann,  ArXiv:1101.1811;
               A. Schilling, H. Grundmann, and R. Dell'Amore, ArXiv:1107.4335.

\eb

\end{document}